\documentclass[sigplan, 10pt, dvipsnames, table]{acmart}
\renewcommand\footnotetextcopyrightpermission[1]{} % removes footnote with conference information in first column
\AtBeginDocument{%
  \providecommand\BibTeX{{%
    Bib\TeX}}}

\setcopyright{none}
 \acmConference[]{SERIAL'22}{Nov. 2022}{Quebec}
\usepackage{algorithmic}
\usepackage{graphicx}
\usepackage{textcomp}
\newcommand*{\myred}{\cellcolor{red!20}}
\newcommand*{\mygreen}{\cellcolor{green!20}}
\newcommand*{\myyellow}{\cellcolor{yellow!20}}
\newcommand*{\myorange}{\cellcolor{orange!20}}
\newcommand*{\mycolor}{\cellcolor{white!40}}
\def\BibTeX{{\rm B\kern-.05em{\sc i\kern-.025em b}\kern-.08em
    T\kern-.1667em\lower.7ex\hbox{E}\kern-.125emX}}
\usepackage{fontawesome}

\usepackage[english]{babel}
\usepackage[utf8]{inputenc}
\usepackage{listings}
\usepackage[frozencache=true,cachedir=.]{minted}
\usepackage{subcaption}
\usepackage{pgfplots}
\pgfplotsset{width=\columnwidth,compat=1.9}
\usepgfplotslibrary{external}
\usetikzlibrary{patterns}
\usepackage{comment}
\usepackage{mathtools}
\usepackage[linesnumbered,vlined,boxed,commentsnumbered, ruled]{algorithm2e} %for psuedo code

\usepackage{enumitem}
\usepackage{yfonts}
\usepackage{textpos}
\usepackage{tikz}
\usepackage{copyrightbox}
\usepackage{color, colortbl}
\usepackage{adjustbox}
\usepackage{comment}
\usepackage{hhline}
\usepackage{caption}

\usepackage{pifont}% http://ctan.org/pkg/pifont
\newcommand{\cmark}{\ding{51}}%
\newcommand{\xmark}{\ding{55}}%

\usepackage{todonotes}

\newcommand{\myparagraph}[1] { \vspace{0.05cm}\textit{#1.}}

\TPMargin{5pt}
\textblockcolour{yellow!20}

\usepackage[most]{tcolorbox}

\begin{document}

\title{Simulating BFT Protocol Implementations at Scale}


\author{Christian Berger}
\email{cb@sec.uni-passau.de}
\orcid{0000-0003-2754-9530}
\affiliation{%
  \institution{University of Passau}
  %\streetaddress{P.O. Box 1212}
  \city{Passau}
  %\state{Bavaria}
  \country{Germany}
}

\author{Sadok Ben Toumia}
\email{bentou01@ads.uni-passau.de}
%\orcid{0000-0003-2754-9530}
\affiliation{%
  \institution{University of Passau}
  %\streetaddress{P.O. Box 1212}
  \city{Passau}
  %\state{Bavaria}
  \country{Germany}
}

\author{Hans P. Reiser}
\email{hansr@ru.is}
\orcid{0000-0002-2815-5747}
\affiliation{%
  \institution{Reykjavík University}
  %\streetaddress{P.O. Box 1212}
  \city{Reykjavík}
  %\state{Bavaria}
  \country{Iceland}
}

\renewcommand{\shortauthors}{Berger et al.}


\begin{abstract}
%With the increasing adoption of blockchain technology, scalable and world-spanning, Byzantine fault-tolerant (BFT) state machine replication (SMR) is getting practical and necessary. 
%Lately, 
%many recent research efforts

The novel blockchain generation of Byzantine fault-tolerant (BFT) state machine replication (SMR) protocols
focuses on scalability and performance to meet requirements of distributed ledger technology (DLT), e.g., decentralization and geographic dispersion.
%, and thus adopts novel ideas to optimize, e.g.,  message dissemination or  
%choosing message 
%aggregation of BFT protocols.
%techniques, or employing scalable cryptographic primitives. 
%Overall, the design space of BFT SMR became even larger and thus more difficult to conquer.
Validating scalability and performance of BFT protocol implementations requires careful evaluation. % to ascertain that goals are met, and, to possibly identify situations in which the system may be bottlenecked, e.g., in terms of network bandwidth or message dissemination time.
While experiments with real protocol deployments usually offer
the best realism, they are costly and time-consuming.
In this paper, we explore simulation of unmodified BFT protocol implementations as 
as a method for cheap and rapid protocol evaluation: We can accurately forecast the performance of a BFT protocol while experimentally scaling its environment, i.e.,  by varying the number of nodes or geographic dispersion.
%explore simulation as modeling technique to make accurate predictions while
%resent a methodology to model executions of real BFT SMR implementations by 
%experimentally scaling the environments of BFT protocol deployments, e.g., by varying the number of
%nodes, their network characteristics as well as their geographic dispersion.
%We present a toolchain that automates and runs large-scale simulations of BFT protocols 
%on top of the existing Phantom simulator. 
Our approach is resource-friendly and preserves application-realism, since existing BFT frameworks can be simply plugged into the simulation engine without requiring code modifications or re-implementation.

\end{abstract}

\begin{CCSXML}
<ccs2012>
   <concept>
       <concept_id>10002944.10011123.10011674</concept_id>
       <concept_desc>General and reference~Performance</concept_desc>
       <concept_significance>500</concept_significance>
       </concept>
   <concept>
       <concept_id>10002944.10011123.10011130</concept_id>
       <concept_desc>General and reference~Evaluation</concept_desc>
       <concept_significance>500</concept_significance>
       </concept>
   <concept>
       <concept_id>10002944.10011123.10011131</concept_id>
       <concept_desc>General and reference~Experimentation</concept_desc>
       <concept_significance>500</concept_significance>
       </concept>
   <concept>
       <concept_id>10010147.10010919.10010172</concept_id>
       <concept_desc>Computing methodologies~Distributed algorithms</concept_desc>
       <concept_significance>500</concept_significance>
       </concept>
 </ccs2012>
\end{CCSXML}

\ccsdesc[500]{General and reference~Performance}
\ccsdesc[500]{General and reference~Evaluation}
\ccsdesc[500]{General and reference~Experimentation}
\ccsdesc[500]{Computing methodologies~Distributed algorithms}
\keywords{simulation, emulation, Byzantine fault tolerance, state machine replication, consensus, Shadow, Phantom}

\maketitle
\pagestyle{plain}

\section{Introduction}
\label{intro}

The current transition towards Web3 presents many challenges in terms of scalability and  performance of distributed ledger technology (DLT). Proof-of-Work~\cite{nakamoto2008bitcoin} is still widely used today, even if it is not environmental sustainable and can often not meet performance requirements of applications~\cite{bonneau2015sok}.
Consequently, coordination-based Byzantine fault-tolerant (BFT) state machine replication (SMR) algorithms experienced renewed research interest~\cite{vukolic2015quest, berger2018scaling} -- resulting in many novel BFT protocols with focus on improving 
scalability~\cite{yin2018hotstuff, crain2021red, cason2021design, neiheiser2021kauri, stathakopoulou2019mir}, or boosting performance under geographic dispersion~\cite{li2020gosig, sousa2015separating, berger2020aware, bonniot2020pnyxdb}. 
%\sketch{To be continued}
% It is challenging to reason about the behvaior of novel BFT protocols. This often requires large-scale evaluations on cloud platforms like AWS, with experiments that employ over several hundred nodes. 
% Evaluations using real protocol deployments offer usually the best realism but are costly and time-consuming
% 

It is a challenging endeavour to reason about the performance and run-time behavior of these novel BFT protocols.
%%% I guess you meant "thorough" instead of "throughout" here??
In  fact, analyzing BFT protocols requires thorough evaluation, which is why 
the research papers describing these protocols contain evaluations with large-scale deployments that are conducted on cloud platforms like AWS, where experiments deploy up to several hundred nodes (e.g., like  in~\cite{yin2018hotstuff, li2020gosig, cason2021design, crain2021red, neiheiser2021kauri} and many more) to demonstrate a protocol's performance and scalability. 
Evaluations using real protocol deployments usually offer the best realism, but are costly and time-consuming. 
Thus, a reasonable alternative for cheap and rapid validation of BFT protocol implementations (that are possibly still in development stage)  % or complements to the accurate but expensive real protocol deployments. 
can be to rely on either emulation or simulation.

% \begin{figure}
%     \centering
%     \includegraphics[width=1\columnwidth]{fig/simulationPaper-intro.pdf}
%     \caption{Real BFT protocol deployments vs. emulation vs. simulation.}
%     \label{fig:intro}
% \end{figure}

\myparagraph{Emulation vs. Simulation}
Emulation tries to duplicate the exact behavior of what is being emulated. 
A clear advantage of emulation is how it preserves realism: BFT protocols still operate in real time and use real kernel and network protocols. As examples serve Mininet~\cite{lantz2010network, handigol2012reproducible}, which creates a realistic virtual network running real kernel, switch and application code on a single machine, or Kollaps~\cite{gouveia2020kollaps}, a decentralized and dynamic topology emulator. 

In contrast to emulation, simulation decouples simulated time from real time and employs abstractions that help accelerate executions: Aspects of interest are captured through a model, which means the simulation only mimics the protocol's environment or its behavior. This has the advantage of easier experimental control, excellent reproducibility (i.e., deterministic protocol runs) and increased scalability when compared to emulation. As potential drawback remains the question of application-realism since the model may not fairly enough reflect reality. Examples of simulators include   ns-3~\cite{riley2010ns} or Shadow~\cite{jansen2012shadow}, which are both discrete-event network simulators for internet applications.

%- - - - - - Contributions - -  - - \\
%In this paper, we first define requirements of an ideal simulator in the context of
%BFT SMR protocols research.
%Subsequently, we survey and compare
%state-of-the-art simulation and emulation tools like ns-3, Mininet and Shadow.

%We conclude that the Shadow simulator, albeit still experiencing some problems, seems to be one of the most suitable candidates.

%Subsequently we explore the possibility of using
%Shadow, a discrete-event
%simulator, in the context of BFT SMR algorithms research. 

%Further, we develop a toolchain
%to automatically generate realistic deployment environments and 
%configure BFT SMR by replicating configurations used in BFT literature for large-scale deployments of interest and
%measure, and plot throughput and latency of such simulated protocol runs.
%Lastly, we discuss and validate obtained results against advertised performance measurements for well-known
%BFT SMR algorithms.

\myparagraph{Evaluating BFT Protocols}
BFTSim~\cite{singh2008bft} is the first simulator that was developed for an eye-to-eye comparison of BFT protocols but it lacks the necessary scalability to be useful for the newer ``blockchain generation'' of BFT protocols (and apparently only up to $n=32$ PBFT~\cite{castro1999practical} replicas can be successfully simulated~\cite{wang2022tool}). A more recent tool~\cite{wang2022tool} allows for scalable simulation of BFT protocols  but it unfortunately requires a complete re-implementation of the BFT protocol in JavaScript. It also can not make predictions on system throughput. Kollaps~\cite{gouveia2020kollaps} was used to reproduce AWS-deployed experiments with BFT-SMaRt~\cite{bessani2014state} and WHEAT~\cite{sousa2015separating} but it is not sufficiently resource-friendly as it executes the real application code in real-time, thus requiring many physical machines to conduct large-scale experiments.

%At a first glance, the possibilities for simulating BFT protocols seem to be only a few, and none of them might be a perfect-fit for performance evaluation at a larger scale.
{ \footnotesize
\begin{table*}[t]
\arrayrulecolor{black}
%\doublerulesepcolor{black!50}
\centering
%\sffamily
\captionsetup{skip=0.5ex}
\begin{tabular}{||*{8}{c||}}
\hhline{|t:=:t:=:t:=:t:=:t:=:t:=t:=|}
%\rowcolor{blue!10}
 & BFTSim~\cite{singh2008bft}  &  BFT Simulator~\cite{wang2022tool} & Kollaps~\cite{gouveia2020kollaps}  &  ns-3~\cite{riley2010ns} &  Mininet~\cite{lantz2010network, handigol2012reproducible} &  \ Phantom~\cite{jansen2022co}  \\ \hhline{|:=::=::=::=::=::=::=:|}
 \mycolor {application layer realism}  & \myred \xmark  & \myred \xmark   & \mygreen \cmark     & \myred \xmark     & \mygreen \cmark         & \mygreen \cmark    \\
\hhline{|:=::=::=::=::=::=::=:|}
\mycolor {realistic networking}  &  \mygreen \cmark & \myyellow  {(\scriptsize \textit{high level model)}}  & \mygreen \cmark     & \mygreen \cmark      & \mygreen \cmark           & \mygreen \cmark      \\
\hhline{|:=::=::=::=::=::=::=:|}
\mycolor {scalability}    & \myred \xmark  & \mygreen \cmark  & \mygreen \cmark     & \mygreen \cmark      & \myred \xmark         & \mygreen \cmark       \\
\hhline{|:=::=::=::=::=::=::=:|}
\mycolor{resource friendliness}    & \mygreen \cmark   & \mygreen \cmark  & \myred \xmark     & \mygreen \cmark      & \myred \xmark          & \mygreen \cmark\\
\hhline{|:=::=::=::=::=::=::=:|}
\mycolor{Byzantine attacker}    & \myorange {(\scriptsize \textit{only bengin faults)}}    &   \mygreen \cmark &  \myred \xmark   & \myred \xmark    &   \myred \xmark      & \myred \xmark       \\
\hhline{|b:=:b:=:b:=:b:=:b:=:b:=:b:=|}
\end{tabular}
\caption{Comparison of different emulators and simulators in the context of BFT protocol research.}
\label{table:comparison}
\end{table*}
}

%\myparagraph{Research Questions}
%In this paper, we explore simulation as a method to evaluate real BFT SMR protocol implementations, which leads us to the following two research questions:

\myparagraph{Research Questions \& Contributions}
We explore simulation as a method to evaluate BFT protocol implementations, which leads us to the following two research questions: 
\begin{itemize}
   \item[\textbf{R1}] What are properties of an ideal %simulation tool for evaluating 
   performance evaluation tool 
   for the "{blockchain} generation" of BFT protocols? 
   \item[\textbf{R2}] Can simulations help us to reason about the behavior of real BFT protocol implementations at larger scale?
\end{itemize}

 Our contributions aim for supporting validations of novel BFT protocol implementations for their practical deployments in large-scale DLT systems. In the following, we summarize our main findings: %We made the following findings: 
%In a search for answers to these research questions, we make the following contributions in this paper:
\begin{itemize}[leftmargin=*]

\item We first compare existing simulators and emulators to analyze properties of an ideal evaluation tool in the context of BFT protocol research. A key finding is, that the state-of-the art is deficient as  there is no resource-friendly evaluation tool to predict the performance (i.e., latency \textit{and} throughput) of BFT protocols at a larger scale. 

    %\item We compare show how to perform scalable simulations for any BFT protocols by plugging-in  select the best for specific deployments (How well do selected protocols in specific deployments?)
    \item We present a tool that automates large-scale simulations of unmodified BFT protocol implementations through the Phantom simulator~\cite{jansen2022co} given a simple experimental description. For the first time, experiments with \emph{existing BFT protocol implementations} can be effortless setup, configured and fed into a simulation engine (Sects.~\ref{section:preliminaries} and ~\ref{section:toolchain}).
    
    \item We discovered that we can faithfully forecast the  
    performance of BFT protocols because performance eventually becomes network-bound at a larger scale. %and to spot possible bottlenecks. 
    Our evaluations compare results obtained from simulations with measurements of real protocol deployments (Sect.~\ref{section:evaluation}).
    %\item Plug \& Play utility to easily evaluate real implementations of state-of-the-art BFT SMR  protocols
\end{itemize}

%\section{In Search for an Ideal Simulator}
\section{Related Work \& Background}
\label{section:in-search}

BFTSim~\cite{singh2008bft} was the first simulator especially tailored for traditional BFT protocols like PBFT~\cite{castro1999practical} or Zyzzyva~\cite{kotla2007zyzzyva}.  %and Q/U~\cite{abd2005fault}. 
Since these protocols were intended to be used for only small groups of replicas, the limited scalability of the simulator was at that time not an issue. However, it makes BFTSim impractical for the newer BFT protocols. BFTSim demands a BFT protocol to be modeled in the P2 language~\cite{loo2005implementing}, which is somewhat error-prone when considering the complexity of, e.g., PBFT's view change or Zyzzyva's many corner cases.
Although BFTSim allows the simulation of faults, it only considers non-malicious behavior and left the extension to more sophisticated Byzantine attacks for future work. It provides realistic networking using ns-2, and is resource-friendly as it runs on a single machine.

Recently, a BFT simulator was presented by Wang et al.~\cite{wang2022tool} which demonstrated resource-friendliness, high scalability,  and comes with an \textit{attacker module} which includes a pre-defined set of attacks (partitioning, adaptive, rushing). %and it can be easily extended through callback functions that would, i.e., allow an (to be implemented) attacker to access and manipulate messages. 
The simulator does not mimic real network protocols, instead it tries to capture network characteristics in  a high-level model where messages can be delayed by some variable sampled from a (to be defined) Gaussian or Poisson distribution. Like BFTSim, it does not provide application layer realism and demands the re-implementation of a BFT protocol in JavaScript. A further drawback is that it cannot measure system throughput, and is thus not suited for reasoning about system performance.
Further, related work also includes stochastic modelling of BFT protocols~\cite{nischwitz2021bernoulli} and validations of BFT protocols through unit test generation~\cite{bano2022twins}.

There are simulators which are dedicated to blockchain research, such as Shadow-Bitcoin~\cite{miller2015shadow}, 
Bitcoin blockchain simulator~\cite{gervais2016security}, 
BlockSim~\cite{faria2019blocksim}, 
SimBlock~\cite{aoki2019simblock} or  ChainSim~\cite{wang2020chainsim}.
These tools mainly focus on building models that capture the characteristics of Proof-of-Work and thus can not easily be adopted or used for BFT protocol research. 

Further, there are tools to emulate or simulate distributed applications: Mininet~\cite{lantz2010network, handigol2012reproducible} and Kollaps~\cite{gouveia2020kollaps} are emulators that allow to create realistic networks (running real internet protocols) and real application code with time being synchronous with the 
wallclock. Naturally, both approaches provide a high degree of realism, which comes at the cost of being less resource-friendly. Mininet is not scalable, a problem which was addressed later by Maxinet~\cite{wette2014maxinet}, which allows Mininet  emulated networks to spawn over several physical machines. Kollaps is a scalable emulator, but also requires many physical machines for large-scale experiments.
 Moreover, ns-3~\cite{riley2010ns} is a resource-friendly and scalable network simulator, but it requires the development of an application model, and thus does not preserve application layer realism.  %The generic simulators and emulators which were not crafted for BFT research do not consider a global Byzantine attacker.
 
Phantom~\cite{jansen2022co} uses a hybrid emulation/simulation
architecture: It executes real applications as native OS processes,
 co-opting the processes into a high-performance network and kernel simulation and thus can scale
to large system sizes. An advantage of this is that it preserves application layer realism as real BFT protocol implementations are executed. At the same time, it is resource-friendly and runs on a single machine. Through its hybrid architecture, Phantom resides in a sweet-spot between ns-3 (pure simulator) and Mininet (pure emulator): It still provides sufficient application realism for the execution of BFT protocols, but is more resource-friendly and scalable than the emulators are.

 As shown in Table~\ref{table:comparison}, there is no perfect solution for simulating BFT protocols at scale yet. If we require both resource-friendliness and scalability, which we think are necessary characteristics to evaluate scalable BFT protocols in an inexpensive way, then only the BFT Simulator of Wang et al.~\cite{wang2022tool} and Phantom~\cite{jansen2022co} are viable options. Comparing these two, we decided to build our evaluation toolchain on top of Phantom, because it allows to plug and play BFT protocol implementations and can measure system throughput.

\section{Preliminaries: Phantom}
\label{section:preliminaries}
\begin{figure}[t]
	\centering
	\includegraphics[width=1\columnwidth]{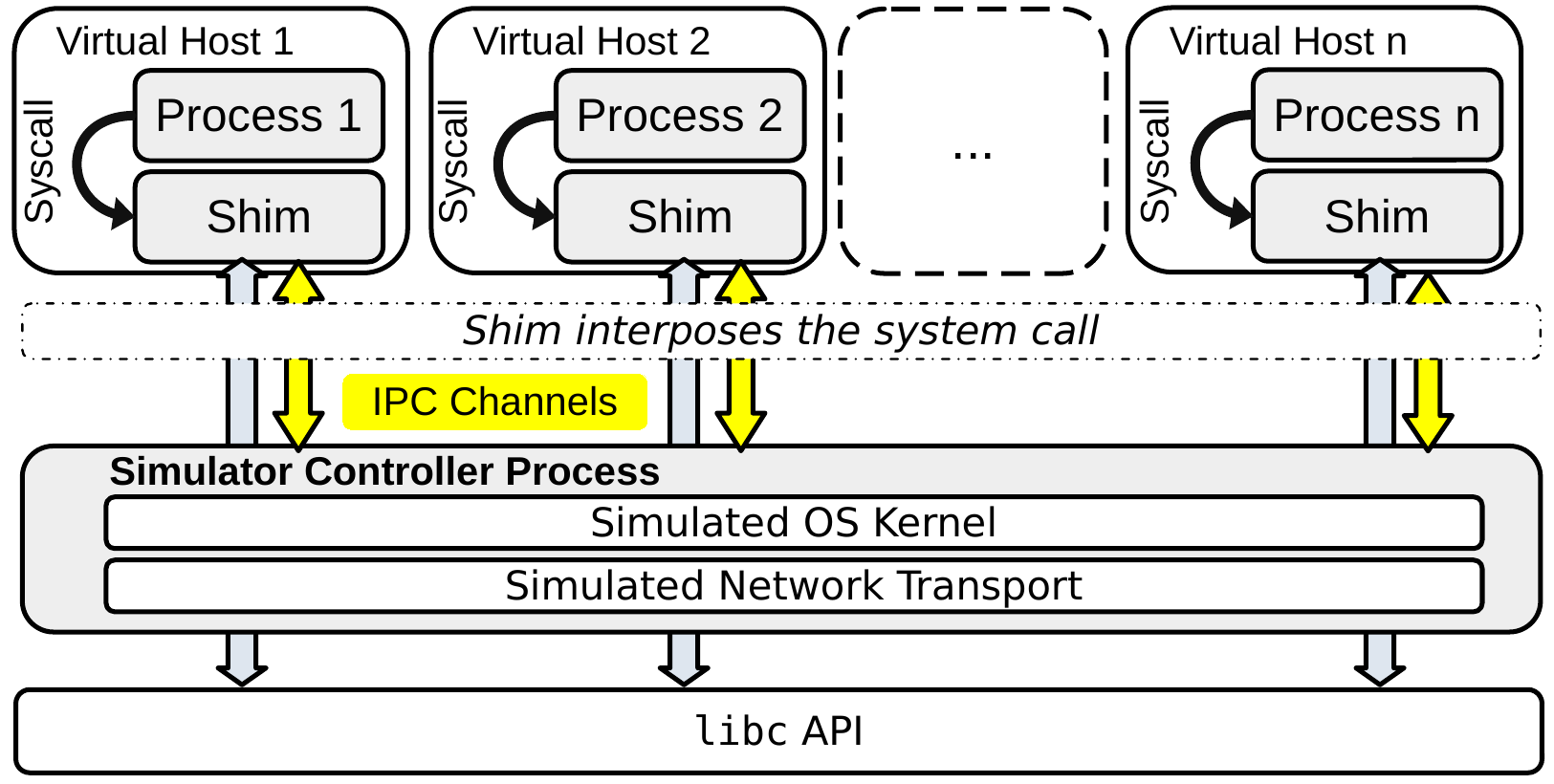}
	\caption{The Phantom architecture (high-level overview).}
	\label{fig:phantom:architecture}
\end{figure}

Phantom uses a hybrid simulation/emulation architecture, in which real, unmodified applications execute as normal processes on Linux and are hooked into the simulation through a system call interface using standard kernel facilities~\cite{jansen2022co}.

In Phantom, a network topology (the \textit{environment}) can be described by specifying a graph, where \textit{virtual hosts} are nodes and communication links are edges. The graph is  attributed: For instance, virtual hosts specify available uplink/downlink bandwidth and links specify latency and packet loss. 
Each virtual host can be used to run one or more applications. This results in the creation of real Linux processes that are initialized by the simulator controller process as managed processes (managed by a Phantom worker). The Phantom worker uses \texttt{LD\_PRELOAD} to preload a shared library (called the \textit{shim}) for co-opting its managed processes into the simulation (see Figure~\ref{fig:phantom:architecture}). \texttt{LD\_PRELOAD} is extended by a second interception strategy, which uses \texttt{seccomp} for cases in which preloading does not work~\cite{jansen2022co}.

The shim constructs an inter-process communication channel (IPC) to the simulator controller process and intercepts functions at the system call interface. While the shim may directly emulate a few system calls,  most system calls are forwarded and handled by the simulator controller process, which simulates kernel and networking functionality (for example the passage of time, I/O operations on \texttt{file}, \texttt{socket}, \texttt{pipe}, \texttt{timer}, event descriptors and packet transmissions)~\cite{jansen2022co}.

%Throughout the simulation, Phantom preserves determinism: It employs a pseudo-random generator, which is seeded from a configuration file to emulate all randomness needed during simulation, in particular the emulation of \texttt{getrandom} or reads of \texttt{/dev/*random}. Each Phantom worker only allows a single thread of execution across all processes it manages so that each of the remaining managed processes/threads are idle, thus preventing concurrent access of managed processes’ memory~\cite{jansen2022co}.
%We briefly display the architecture in Figure~\ref{fig:phantom:architecture}, for more details we refer the reader to the Phantom paper~\cite{jansen2022co}.

\begin{figure}[t]
    \centering
    \includegraphics[width=0.83\columnwidth]{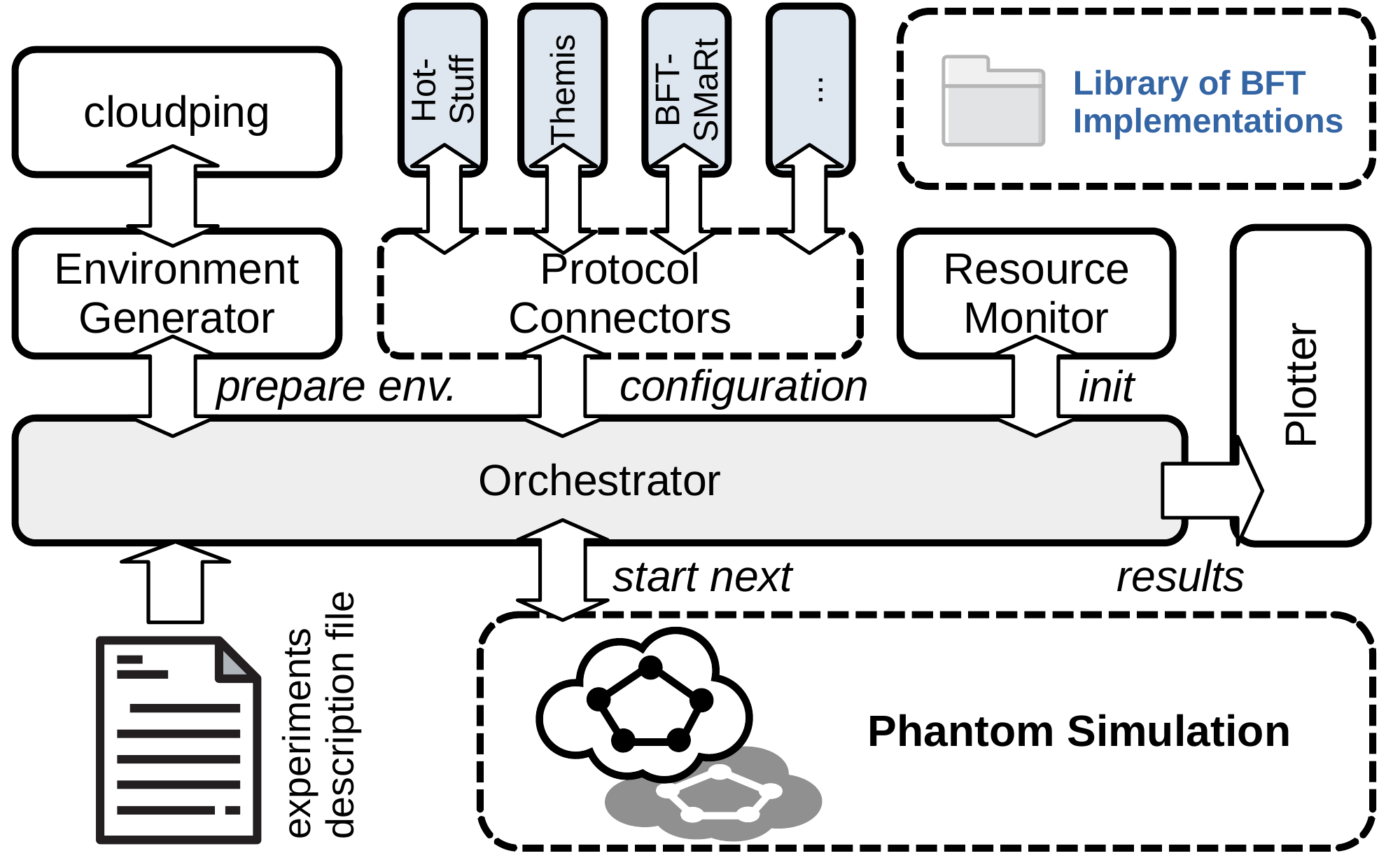}
    \caption{Our toolchain architecture for automating the setup of simulation runs of BFT protocols with Phantom.}
    \label{fig:toolset:architecture}
\end{figure}
\section{A Simulation Toolchain for BFT}
\label{section:toolchain}

Large-scale simulations of BFT protocols with Phantom requires additional tooling support. This is mainly because of the following reasons: 
%\subsection{Necessity of Tooling Support}
First, Phantom requires to generate realistic and large network topologies for an arbitrary system size  and the characteristics of their communication links should ideally resemble real-world deployments. This is crucial to allow realistic simulation of wide-area network environments. 
Second, we need aid in setting up the BFT protocol implementations for their deployment, since 
bootstrapping a BFT protocol in Phantom involves many steps that can be tedious, error-prone and protocol-specific.
This means, for instance, the generation of protocol-specific run-time artifacts like cryptographic key material, or configuration files which differ for every BFT protocol. 
Third, in the process of developing and testing BFT algorithms, different combinations of protocol settings result in numerous experiments to be conducted. Since Phantom simulations run in virtual time, they can take hours, depending on the host system’s specifications. For the sake of user experience and convenience, we find it is necessary for experiments to be specified in bulk and ran sequentially without any need for user-intervention.
Fourth, we may want to track and evaluate resources needed during simulation runs, such as CPU utilization and memory usage. 
Fifth, when Phantom produces results, they are resided in the file system and for convenience we want to aggregate  measurements of several simulations and map these to diagrams displaying to-be-specified metrics like throughput or latency.

 These reasons led us to develop Delphi-BFT\footnote{Code open-source available at \url{https://github.com/Delphi-BFT/tool}.}, a tool on top of Phantom to simplify and accelerate the evaluation of unmodified BFT protocol implementations.

%\subsection{Selection of BFT Consensus Protocols}
%\subsection{Preparation: Capturing CPU Utilization}
%\subsection{Modelling and Scaling Environments}
%\subsection{Protocol Configuration and Parameterization}
%\subsection{Measurements}

\subsection*{Architecture}
Delphi-BFT is composed of several components (see Fig~\ref{fig:toolset:architecture}) and follows a modular architecture, in that it is not tailored
to a specific BFT protocol, but is easily extensible. %Further, different components be easily replaced, provided the user implements specific
%functions.

\myparagraph{Orchestrator} The toolchain is administered by an orchestrator that manages all tools,
i.e., for preparing an environment, configuring runtime artifacts for a BFT protocol, and initializing a resource monitor. The orchestrator invokes protocol connectors to set up a BFT protocol and loads \textit{experiments description files} which contain a set of experiments to be conducted for the specified BFT protocol. Finally, it starts Phantom, once an experiment is ready for its execution. 

\myparagraph{Environment Generator}
The environment generator creates network topologies as complete graph for any  system size. The network topologies  resemble realistic deployment scenarios for a LAN or WAN setting. To create network graphs with network links reflecting a realistic geographic dispersion of nodes, the environment generator employs a cloudping component, which retrieves real round-trip latencies between all AWS regions from Cloudping\footnote{See \url{https://www.cloudping.co/grid}.}. This allows the tool to create network topologies which resemble real BFT protocol deployments on the AWS cloud infrastructure.

\myparagraph{Protocol Connectors} For each BFT protocol implementation that we want to simulate, it is necessary to create protocol configuration files and necessary keys. Since protocol options and cryptographic primitives vary depending on the concrete BFT protocol, we implement the protocol-specific setup routine as a tool called protocol connector, which is invoked by the orchestrator.
A connector must implement the methods \texttt{build()} and \texttt{configure()}. 
This way, it is simple to extend our toolchain and support  new BFT protocols, as it only requires writing a new protocol connector (in our experience this means writing between 100 and 200 LoC).

\myparagraph{Resource Monitor} The orchestrator initializes a resource monitor to collect information on resource consumption (like allocated memory and CPU time) during simulation runs and also the total simulation time. The user can use these statistics as indicators towards a possible need for vertically scaling the host machine and as rough estimates for the necessary resources to run larger simulations.

\myparagraph{Plotter} Results are stored to the file system by Phantom. They can be aggregated and mapped to specific diagrams for specifiable metrics like latency of throughput. For instance, it can create diagrams that display the performance of a BFT protocol for increasing system scale which aggregate several simulation runs for an increasing $n$ (or any other variable).
%\sketch{\textit{cloudping.js} to sample real AWS latency data between all regions from Cloudping}
%\sketch{\textit{environment-generator.js} creates Shadow complete-graph network topologies for any scale $n$ with either fixed latency (for LAN experiments) or retrieved from cloudping (for WAN experiments)}
%\sketch{\textit{protocol connectors}: for every target protocol, set up e.g., keys and create configuration file for deployment in the virtual Shadow topology}
%\sketch{\textit{resource-monitor.js}: Collects information on resource consumption during a simulation}
%\sketch{\textit{orchestrator}: manages all above tools and, once all artifacts (config files, network topology) are ready, starts the simulation}
%\sketch{\textit{plot.js}: Plot results obtained from several simulation runs}

\begin{table}[t]
	\centering
	\resizebox{\columnwidth}{!}{%
		\begin{tabular}{llll}
			\hline
			\textbf{framework}   & \textbf{BFT protocol} & \textbf{language} & \textbf{repo on github.com}  \\ \hline
			libhotstuff~\cite{yin2018hotstuff} & Hot-Stuff    & C++      & /hot-stuff/libhotstuff \\
			themis~\cite{rusch2019themis}      & PBFT         & Rust     & /ibr-ds/themis         \\
			bft-smart~\cite{bessani2014state}   & BFT-SMaRt    & Java     & /bft-smart/library     \\ \hline
		\end{tabular}%
	}
	\caption{BFT protocols that we employed for our evaluation.}
	\label{table:BFT-frameworks}
\end{table}
\section{Results}
\label{section:evaluation}

In this section, we try to find an answer to our second research question.
%conduct a prelimiary evaluation of our approach.
We first use results from the HotStuff paper~\cite{yin2018hotstuff} to compare our simulation results with real measurements and reason about resource utilization of our simulations. Then, we show that we can achieve realistic results under geographic dispersion by simulating a WAN topology for BFT-SMaRt~\cite{bessani2014state} replicas. Further, we experiment with a Rust-based PBFT implementation~\cite{rusch2019themis} to demonstrate  compatibility with a third programming language (see Table~\ref{table:BFT-frameworks}).

% \usepackage{graphicx}

%\myparagraph{Exploring the Network Bound of HotStuff} 
\subsection{HotStuff at Increasing System Scale} In our first evaluation, we try to mimic the evaluation setup of the HotStuff paper~\cite{yin2018hotstuff} to compare their measurements with our simulation results. Their setup consists of more than hundred virtual machines deployed in an AWS data center; each machine has up to 1.2 GB/s bandwidth and  there is less than 1 ms latency between each pair of machines (we use 1~ms in the simulation). 
The employed batch size is $400$. We compare against two measurement series: "p1024" where the payload size of request and responses is 1024 bytes and  
 "10ms" with empty payload, but the latency of all communication links is set to $10$~ms.
Our goal is to investigate how faithfully the performance of HotStuff can be predicted by regarding only  the networking capabilities of replicas, which manifests at the point where the network becomes the bottleneck for system performance.
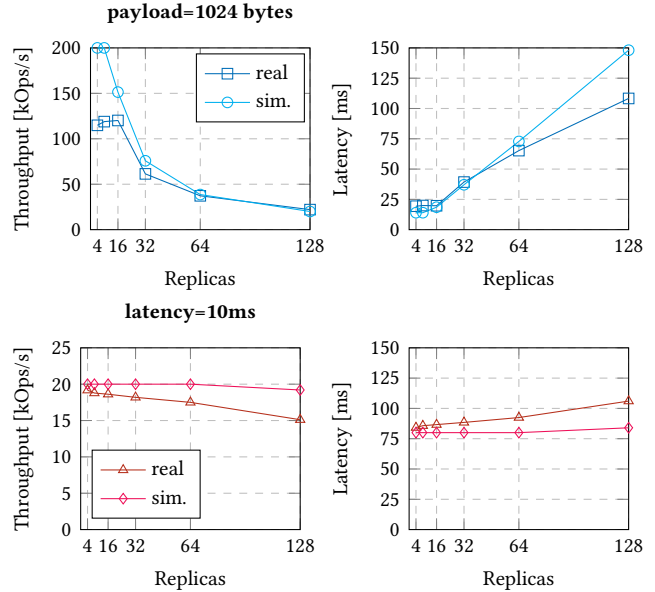
\begin{figure}[t]
	\centering
	%   \begin{subfigure}[b]{\columnwidth}
		\begin{subfigure}[b]{.49\columnwidth}
			\centering
			 \begin{tikzpicture}
    \begin{axis}[
%width= 8cm,
%height=5cm,
%font= \footnotesize, %\footnotesize,
width= 4.5cm,
height=4cm,
font= \footnotesize,
    title={\textbf{payload=1024 bytes}},
    xlabel={Replicas},
    ylabel={Throughput [kOps/s]},
    xmin=0, xmax=128,
    ymin=0, ymax=200,
    xtick={ 4, 16, 32, 64, 128},
    ytick={0, 50, 100,150,200,250, 300},
   % legend pos=south east,
   % legend columns = 2,
    %legend style={at={(0.98, 0.8)}},
    %legend style={at={(1, 1.05)}, draw=none},
    legend style={at={(0.48,0.78)},anchor=west, legend columns=1},
    legend cell align={left},
    ymajorgrids=true,
    xmajorgrids=true,
    grid style=dashed,
]

\addplot[
    color=NavyBlue,
    mark=square,
    ]
    table [x=replicas,y=throughput] {data/throughput-HS3-p1024.txt};
     \addplot[
    color=ProcessBlue,
    mark=o,
    ]
     table [x=replicas,y=throughput] {data/phantom-thr-hs3-p1024-1ms.txt};
% \addplot[
%     color=BrickRed,
%     mark=triangle,
%     ]
%     table [x=replicas,y=throughput] {data/throughput-HS3-L10.txt};

%  \addplot[
%     color=OrangeRed,
%     mark=diamond,
%     ]
%      table [x=replicas,y=throughput] {data/phantom-thr-hs3-p0-10ms.txt};
   
   \legend{real, sim.}
\end{axis}
\end{tikzpicture} % - avoid white space
\vskip -0.1 cm
			% \caption{Throughput.}
			\label{fig:hs3:throughput}
		\end{subfigure}
		% \begin{subfigure}[b]{\columnwidth}
			\begin{subfigure}[b]{.49\columnwidth}
				\centering
				 \begin{tikzpicture}
    \begin{axis}[
width= 4.5cm,
height=4cm,
font= \footnotesize,
    xlabel={Replicas},
    ylabel={Latency [ms]},
    xmin=0, xmax=128,
    ymin=0, ymax=150,
    xtick={4, 16, 32, 64, 128},
    ytick={0, 25, 50, 75, 100, 125, 150},
    %legend pos=north east,
    legend columns = 3,
   legend style={at={(0.33,1.12)},anchor=west, legend columns=3},
    legend cell align={left},
    ymajorgrids=true,
    xmajorgrids=true,
    grid style=dashed,
]

\addplot[
    color=NavyBlue,
    mark=square,
    ]
    table [x=replicas,y=latency] {data/latency-HS3-p1024.txt};
    
        \addplot[
    color=ProcessBlue,
    mark=o,
    ]
    table [x=replicas,y=latency] {data/phantom-lat-hs3-p1024-1ms.txt};
   %  \addplot[
   % color=BrickRed,
   %  mark=triangle,
   %  ]
   %  table [x=replicas,y=latency] {data/latency-HS3-L10.txt};

   %   \addplot[
   %  color=OrangeRed,
   %  mark=diamond,
   %  ]
   %   table [x=replicas,y=latency] {data/phantom-lat-hs3-p0-10ms.txt};
    
% \legend{HS3-p1024, HS3-10ms, P}
\end{axis}
\end{tikzpicture} % - avoid white space
\vskip -0.1 cm
				%   \caption{Latency.}
				\label{fig:hs3:latency}
			\end{subfigure}
			\begin{subfigure}[b]{.49\columnwidth}
				\centering
				 \begin{tikzpicture}
    \begin{axis}[
%width= 8cm,
%height=5cm,
%font= \footnotesize, %\footnotesize,
width= 4.5cm,
height=4cm,
font= \footnotesize,
   title={\textbf{latency=10ms}},
    xlabel={Replicas},
    ylabel={Throughput [kOps/s]},
    xmin=0, xmax=128,
    ymin=0, ymax=25,
    xtick={ 4, 16, 32, 64, 128},
    ytick={0, 5, 10, 15, 20, 25},
   % legend pos=south east,
   % legend columns = 2,
    %legend style={at={(0.98, 0.8)}},
    %legend style={at={(1, 1.05)}, draw=none},
    legend style={at={(0.05, 0.25)},anchor=west, legend columns=1},
    legend cell align={left},
    ymajorgrids=true,
    xmajorgrids=true,
    grid style=dashed,
]

\addplot[
    color=BrickRed,
    mark=triangle,
    ]
    table [x=replicas,y=throughput] {data/throughput-HS3-L10.txt};

 \addplot[
    color=OrangeRed,
    mark=diamond,
    ]
     table [x=replicas,y=throughput] {data/phantom-thr-hs3-p0-10ms.txt};
   
  \legend{real, sim.}
\end{axis}
\end{tikzpicture} % - avoid white space
\vskip -0.1 cm
				% \caption{Throughput.}
				\label{fig:hs3:throughput}
			\end{subfigure}
			% \begin{subfigure}[b]{\columnwidth}
				\begin{subfigure}[b]{.49\columnwidth}
					\centering
					 \begin{tikzpicture}
    \begin{axis}[
width= 4.5cm,
height=4cm,
font= \footnotesize,
    xlabel={Replicas},
    ylabel={Latency [ms]},
    xmin=0, xmax=128,
    ymin=0, ymax=150,
    xtick={4, 16, 32, 64, 128},
    ytick={0, 25, 50, 75, 100, 125, 150},
    %legend pos=north east,
    legend columns = 3,
   legend style={at={(0.33,1.12)},anchor=west, legend columns=3},
    legend cell align={left},
    ymajorgrids=true,
    xmajorgrids=true,
    grid style=dashed,
]
    table [x=replicas,y=latency] {data/phantom-lat-hs3-p1024-1ms.txt};
    \addplot[
   color=BrickRed,
    mark=triangle,
    ]
    table [x=replicas,y=latency] {data/latency-HS3-L10.txt};

     \addplot[
    color=OrangeRed,
    mark=diamond,
    ]
     table [x=replicas,y=latency] {data/phantom-lat-hs3-p0-10ms.txt};
    
% \legend{HS3-p1024, HS3-10ms, P}
\end{axis}
\end{tikzpicture} % - avoid white space
\vskip -0.1 cm
					%   \caption{Latency.}
					\label{fig:hs3:latency}
				\end{subfigure}
				\caption{Performance of HotStuff and Simulated-HotStuff.}
				\label{fig:hs3}
			\end{figure}

\textit{Observations}. We display our results in Figure~\ref{fig:hs3}. The simulation results for the payload experiment indicate a similar trend as the real measurements, where performance starts to drop for $n \geq 32$. For a small sized replica group, the network simulation predicts higher performance: 200k tx/s. This equals the theoretical maximum limited only through the 1 ms link latency which leads to pipelined HotStuff committing a batch of 400 requests every 2 ms.
The difference in throughput decreases once the performance of HotStuff becomes more bandwidth-throttled (at $n\geq 32$). We also achieve close results in the "10ms" setting: 80 ms in the simulation vs 84.1 ms real, and 20k tx/s in the simulation vs. 19.2k tx/s real for $n=4$; but with an increasing difference for higher $n$, i.e., 84 ms vs. 106 ms and 19k.2 tx/s vs. 15.1k tx/s for $n=128$.

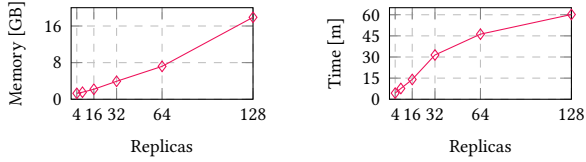
\begin{figure}[t]
	\centering
	\begin{subfigure}[b]{.49\columnwidth}
		\centering
		{ 
\begin{tikzpicture}
    \begin{axis}[
width= 4cm,
height=2.8cm,
font= \scriptsize,
    xlabel={Replicas},
    ylabel={ Memory [GB]},
    xmin=0, xmax=128,
    ymin=0, ymax=20,
    xtick={4, 16, 32, 64, 128},
    ytick={0, 8, 16},
    legend pos=south east,
    legend columns = 2,
    legend style={at={(0.8, 0.8)}},
    legend cell align={left},
    ymajorgrids=true,
    xmajorgrids=true,
    grid style=dashed,
]

\addplot[
    color=OrangeRed,
    mark=diamond,
    ]
    table [x=replicas,y=libhotstuff] {data/resource-consumption-memory.txt};
  % \legend{libhotstuff}
\end{axis}
\end{tikzpicture} % - avoid white space
\vskip -0.1 cm

}
		%   \caption{Allocated memory.}
		\label{fig:resources:memory}
	\end{subfigure}
	\begin{subfigure}[b]{.49\columnwidth}
		\centering
		 \begin{tikzpicture}
    \begin{axis}[
width= 4cm,
height=2.8cm,
font= \scriptsize,
    xlabel={Replicas},
    ylabel={Time [m]},
    xmin=0, xmax=128,
    ymin=0, ymax=65,
    xtick={4, 16, 32, 64, 128},
    ytick={0, 15, 30, 45, 60},
    legend pos=south east,
    legend columns = 2,
    legend style={at={(0.98, 0.05)}},
    legend cell align={left},
    ymajorgrids=true,
    xmajorgrids=true,
    grid style=dashed,
]

\addplot[
    color=OrangeRed,
     mark=diamond,
     ]
    table [x=replicas,y=libhotstuff] {data/resource-consumption-time.txt};
   
 %  \legend{libhotstuff}
\end{axis}
\end{tikzpicture} % - avoid white space
\vskip -0.1 cm
		%  \caption{Simulation time.}
		\label{fig:resources:time}
	\end{subfigure}
	\caption{Resource consumption of simulations.}
	\label{fig:resource:consumption}
\end{figure}

\textit{Resource Usage.} Further, we investigate how resource utilization, i.e. memory usage and simulation time, grows with an increasing system scale. We run our HotStuff "10ms" simulations (which display a somewhat steady system performance for increasing system scale) on an Ubuntu 20.04 VM with 48~GB memory and 20 threads (16 threads used for simulation) on a host with an Intel Xeon Gold 6210U CPU at 2.5 GHz. We observe that active host memory and elapsed time grow with increasing system scale (see Fig.~\ref{fig:resource:consumption}). We think it should be feasible to simulate up to 512 HotStuff replicas with a well-equipped host (with e.g., 64 GB RAM).

%Like in the HotStuff paper, 
%We conduct our experiments with up to $n=128$ replicas.

% \sketch{Evaluation 1: Simulation results vs. real measurements from literature.
% \textbf{HS3-p1024}. very low latency (< 1 ms) and look-ahead parameter. High bandwidth  (see Hot-Stuff paper: 1.2 Gigabytes per second (9.6 Gbit)) Request size at 1024 Bytes;  batch size 400\\
% \textbf{HS3-10ms}. higher latency of 10 ms (see Hot-Stuff paper), request size is 0 \\
%  batch size 400\\
 
% ---- \\

% Results: 2 Diagrams: one for Latency, one for Throughput. Each Diagram has 4 data rows: (HotStuff, HotStuff-10ms, Phantom-HotStuff, PhantomHotStuff-10ms;
% }

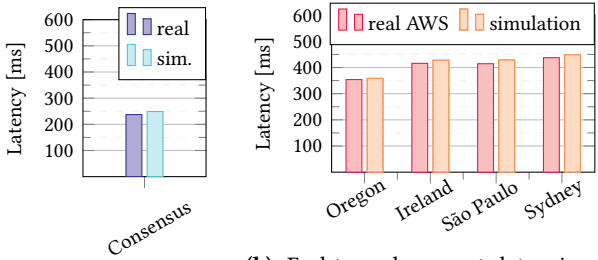
\begin{figure}[t]
    \centering
  \begin{subfigure}[h]{0.38\columnwidth}
 \begin{tikzpicture} 
    \begin{axis}[ 
    font= \footnotesize,
     ylabel={Latency [ms]}, 
     xticklabels from table={data/consensus-latencies.txt}{region},   
        x tick label style={rotate=30,anchor=east,  xshift=15pt, yshift=-14pt,   font= \footnotesize},
     ybar=2pt,  %configures bar shift 
     bar width=6pt,
      % width=0.6\columnwidth, 
    height=3.67cm,
       xtick=data, 
       ytick = {100,200,300,400,500,600},
        ymin=0,
        ymax=600,
        xmin=0.5,
        xmax=1.5,
    ymajorgrids=true,
    yminorgrids=true,
    minor grid style={dashed,gray!10},
    minor tick num=1,
    %legend pos= north west,% Legende oben links in Abb. 
    legend style={at={(1, 1.08)},
    legend columns = 1,
    legend cell align=left% Rechtsbündige Ausrichtung der Legende 
    }
    ] 
      \addplot
      [draw = Blue,
        fill = Blue!30!white]   
        table[ 
          x=regionNr, 
          y=real   
          ] 
      {data/consensus-latencies.txt}; 
     \addlegendentry{real }; 
   
            \addplot 
      [draw = SkyBlue, 
        fill = SkyBlue!30!white ]
        table[ 
          x=regionNr, 
          y=simulated 
          ] 
      {data/consensus-latencies.txt}; 
      \addlegendentry{sim.}; 
    \end{axis}
\end{tikzpicture} 
\vskip -0.1 cm
      \caption{Consensus latency.}
    \label{fig:consensus-latency}
  \end{subfigure}
  \begin{subfigure}[htb]{0.6\columnwidth}
  \begin{tikzpicture} 
    \begin{axis}[ 
    font= \footnotesize,
     ylabel={Latency [ms]}, 
     xticklabels from table={data/client-latencies.txt}{region},   
        x tick label style={rotate=30,anchor=east,  xshift=10pt, yshift=-4pt,   font= \footnotesize},
     ybar=2pt,  %configures bar shift 
     bar width=6pt,
      % width=0.6\columnwidth, 
    height=3.67cm,
       xtick=data, 
       ytick = {100,200,300,400,500,600},
        ymin=0,
        ymax=600,
        xmin=0.5,
        xmax=4.5,
    ymajorgrids=true,
    yminorgrids=true,
    minor grid style={dashed,gray!10},
    minor tick num=1,
    %legend pos= north west,% Legende oben links in Abb. 
    legend style={at={(1, 1.08)},
    legend columns = 5,
    legend cell align=left% Rechtsbündige Ausrichtung der Legende 
    }
    ] 
      \addplot 
      [draw = OrangeRed,
        fill = OrangeRed!30!white]   
        table[ 
          x=regionNr, 
          y=real   
          ] 
      {data/client-latencies.txt}; 
     \addlegendentry{real AWS \  }; 
   
            \addplot 
      [draw = Peach, 
        fill = Peach!30!white,
        postaction={pattern=north east lines,pattern color=Peach!30!white}]   
        table[ 
          x=regionNr, 
          y=simulated 
          ] 
      {data/client-latencies.txt}; 
      \addlegendentry{simulation}; 
    \end{axis}
\end{tikzpicture} 
\vskip -0.1 cm
\caption{End-to-end request latencies observed by clients in AWS regions.}
\label{fig:request-latency}
  \end{subfigure}
         \caption{Comparison of a real BFT-SMaRt WAN deployment on the AWS infrastructure with its simulated counterpart.}
    \label{fig:bftsmart-latencies}
\end{figure}

%\myparagraph{WAN Experiment with BFT-SMaRt}
\subsection{BFT-SMaRt under Geographic Dispersion}
Next, we experiment with geographic dispersion of BFT-SMaRt replicas, where each replica is located in a distinct AWS region. Our experimental setup is thus  similar to experiments found in papers that research on latency improvements~\cite{sousa2015separating,berger2020aware, berger2021making}. We employ a  $n=4$ configuration and choose the regions Oregon, Ireland, São Paulo and Sydney for the deployment of a replica and a client application each. 
We run clients one after another, and each samples 1000 requests without payload and measures end-to-end latency, while the leader replica (in Oregon) measures the system's consensus latency. Further, we create an experiments description file mimicing  this experiment (see Appendix C) and run it through our simulation toolchain to compare our simulation results with real measurements.

\textit{Observations}. We notice that consensus latency is slightly higher in the simulation (237 ms vs. 249 ms), and further, the simulation results also display slightly higher end-to-end request latencies in all clients (see Figure~\ref{fig:bftsmart-latencies}).
The deviation between simulated and real execution is the lowest in Oregon (1.3\%) and the highest in São Paulo (3.5\%).

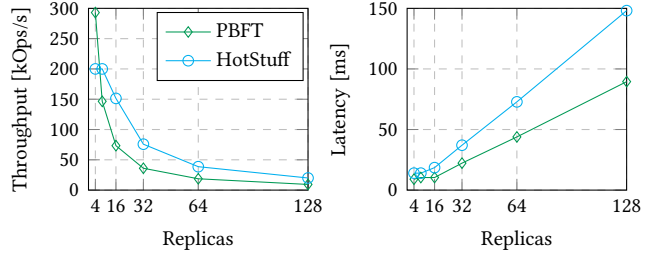
\begin{figure}[t]
	\centering
	%   \begin{subfigure}[b]{\columnwidth}
		\begin{subfigure}[b]{.49\columnwidth}
			\centering
			 \begin{tikzpicture}
    \begin{axis}[
%width= 8cm,
%height=5cm,
%font= \footnotesize, %\footnotesize,
width= 4.5cm,
height=4cm,
font= \footnotesize,
    xlabel={Replicas},
    ylabel={Throughput [kOps/s]},
    xmin=0, xmax=128,
    ymin=0, ymax=300,
    xtick={ 4, 16, 32, 64, 128},
    ytick={0, 50, 100,150,200,250, 300},
    legend pos=south east,
    legend columns = 1,
    legend style={at={(0.98, 0.6)}},
    legend cell align={left},
    ymajorgrids=true,
    xmajorgrids=true,
    grid style=dashed,
]

\addplot[
    color=ForestGreen,
    mark=diamond,
    ]
    table [x=replicas,y=throughput] {data/phantom-thr-pbft-p1024.txt};

\addplot[
    color=ProcessBlue,
    mark=o,
    ]
    table [x=replicas,y=throughput] {data/phantom-thr-hs3-p1024-1ms.txt};
    
% \addplot[
%     color=NavyBlue,
%     mark=square,
%     ]
%     table [x=replicas,y=throughput] {data/throughput-HS3-p1024.txt}; %% ADAPT HERE
% \addplot[
%     color=BrickRed,
%     mark=triangle,
%     ]
%     table [x=replicas,y=throughput] {data/throughput-HS3-L10.txt}; %% ADAPT HERE
   \legend{PBFT, HotStuff}
\end{axis}
\end{tikzpicture} % - avoid white space
\vskip -0.1 cm
			%   \caption{Throughput.}
			\label{fig:pbft:throughput}
		\end{subfigure}
		% \begin{subfigure}[b]{\columnwidth}
			\begin{subfigure}[b]{.49\columnwidth}
				\centering
				 \begin{tikzpicture}
    \begin{axis}[
width= 4.5cm,
height=4cm,
font= \footnotesize,
    xlabel={Replicas},
    ylabel={Latency [ms]},
    xmin=0, xmax=128,
    ymin=0, ymax=150,
    xtick={4, 16, 32, 64, 128},
    ytick={0, 50, 100, 150},
    legend pos=south east,
    legend columns = 2,
    legend style={at={(0.98, 0.8)}},
    legend cell align={left},
    ymajorgrids=true,
    xmajorgrids=true,
    grid style=dashed,
]

 \addplot[
    color=ForestGreen,
    mark=diamond,
    ]
    table [x=replicas,y=latency] {data/phantom-lat-pbft-p1024.txt};

   \addplot[
    color=ProcessBlue,
    mark=o,
    ]
    table [x=replicas,y=latency] {data/phantom-lat-hs3-p1024-1ms.txt};

% \addplot[
%     color=NavyBlue,
%     mark=square,
%     ]
%     table [x=replicas,y=latency] {data/latency-HS3-p1024.txt}; %% ADAPT HERE
    
%     \addplot[
%   color=BrickRed,
%     mark=triangle,
%     ]
%     table [x=replicas,y=latency] {data/latency-HS3-L10.txt}; %% ADAPT HERE
 %  \legend{HS3-p1024, HS3-10$ms$}
\end{axis}
\end{tikzpicture} % - avoid white space
\vskip -0.1 cm
				%  \caption{Latency.}
				\label{fig:pbft:latency}
			\end{subfigure}
			\caption{Simulation results of PBFT vs. HotStuff for 1~KiB.}
			\label{fig:pbft}
		\end{figure}
\subsection{PBFT at Increasing System Scale} We run simulations with 1KiB payload with Themis~\cite{rusch2019themis} (a Rust-based implementation of PBFT) to compare the results against our  HotStuff simulation results.
%\sketch{Repeat the first experiment with Themis/PBFT}

\textit{Observations}. PBFT initially outperfroms HotStuff, but then its throughput decreases more swiftly (as can be seen in the sharper curve in Figure~\ref{fig:pbft}). At $n=128$, PBFT  achieves up to 9.3k tx/s while HotStuff achieves up to 20k tx/s.

% \section{Discussion}
% \label{section:discussion}

% \sketch{I feel that a discussion of results might be important, but we are running out of space}

\section{Future Work}
\label{section:futureWork}
%\sketch{Conclusion comes here}

\myparagraph{Extending Evaluations} For future work, we intend to extend our evaluations to more BFT protocols, in particular, to evaluate the effectiveness of different communication strategies, like Gosig~\cite{li2020gosig} (gossip) or Kauri~\cite{neiheiser2021kauri} (tree-based) and compare them with the results obtained from Hot-Stuff (star-based) and PBFT (clique). In particular, we can explore the performance of these protocols under different network characteristics and for an increasing system scale. A high-level simulation model previously studied the effect of different message exchange patterns of BFT protocols~\cite{silva2020comparison} but it lacks applicability for reasoning about real system metrics.

\myparagraph{CPU Model} We think a CPU model could improve simulation results for evaluations of either (1) small sized replica groups or (2)~experiments with empty payload  -- in both cases the CPU may be the dominating factor and not the network. Up to now, we use Phantom only as a network simulator and all computations, such as creating or verifying signatures, take no time. It might be possible to capture most of the computational work by only modeling a few methods, in particular, the cryptographic primitives (like in BFTSim~\cite{singh2008bft}).
Currently, Phantom plans the introduction of a CPU model as a future milestone for development and we will try to utilize it to improve our simulation results.

\myparagraph{Attacker Model} Moreover, we have in view to introduce an attacker model to reason about the impact of attacks on system performance. % the performance of BFT protocol implementations in the presence of a Byzantine attacker. %At the moment, we can schedule begin (crash-stop) faults by letting faulty replicas' processes terminate during the simulation run.
For this reason, we seek inspiration from the Twins~\cite{bano2022twins} methodology, a recent approach for validating BFT protocols: Twins is an unit test case generator that can simulate  Byzantine attacks by duplicating cryptographic identities of replicas (which then leads to forgotten protocol states, or equivocations) and it can be quite useful in a simulator to explore a variety of attacking scenarios.

%\sketch{Evaluation with more BFT protocols to evaluate different communication topologies, i.e., Gosig (gossip) or Kauri (tree-based)}
%\sketch{CPU Model (currently a future milestone of development see \url{https://github.com/shadow/shadow/milestone/48}}
%\sketch{Attacker Model (get some inspiration from Twins paper?), Twins~\cite{bano2022twins} is a recent approach for validation (unit test case generation) of Byzantine attacks (used to reason about liveness and safety) that can be useful in a simulator.}

% \section*{Acknowledgement}
% This work has been funded by the Deutsche Forschungsgemeinschaft (DFG, German Research Foundation) grant number 446811880 (BFT2Chain).
% We are thankful for the help from the responsive Phantom development team on github.

\begin{acks}
This work has been funded by the Deutsche Forschungsgemeinschaft (DFG, German Research Foundation) grant number 446811880 (BFT2Chain).
We are thankful for the help we received from the Phantom development team.
\end{acks}

\bibliographystyle{ACM-Reference-Format}
\bibliography{bibliography}

\appendix

\section*{Appendix}

In this section, we first define the properties that we used to create our comparison of evaluation tools (Sect. A). Further, we  
 explain how to (B) reproduce this paper's simulation results on your own machine, (C) create your own experiments to run on our toolchain, and (D) connect your own BFT protocol implementation with our toolchain by writing a protocol connector.

\subsection*{\textbf{A} Properties of an Ideal Simulator for BFT Research}

\label{section:properties}

In the following, we briefly explain a set of distilled properties that we employed to create our comparison:

\begin{itemize}[leftmargin=*]
	\item \textbf{Realism}: The simulator allows us to reason about the real protocol behavior fairly enough. We can distinguish this further into the
	characteristics of  \textit{realistic networking} and \textit{application layer realism} (which means the application model of a BFT protocol  matches its implementation).
	\item \textbf{Scalability}: The simulator can handle up to the magnitude of over $10^3$ of nodes executing the BFT protocol and can also handle the geographic dispersion of nodes, i.e., by maintaining a large network topology.
	\item \textbf{Resource friendliness:} To conduct experiments, it is not necessary to have many physical machines at hand.
	\item \textbf{Reproducibility}: Repeated runs give similar results (or even the same results in the case of simulation runs which can be deterministic).
	\item \textbf{Experimental controllability}: It is simple to study isolatable factors, e.g., controlling the environment, or parameters of the protocol.
	\item \textbf{Fault induction / Byzantine attacker:} The simulator provides support to induce faults, for instance, dropping messages, crashing nodes or more complex, in particular malicious attacking behavior orchestrated by an attacker.
	
\end{itemize}

The first five properties are favorable for any simulations of distributed systems. If models use re-implementation or if simplifications are used, then application layer realism is hard to achieve. This is because BFT protocols are generally difficult to implement, and a re-implementation (that may even simplify the protocol) can easily induce bugs (a fact that is also stressed in BFTSim~\cite{singh2008bft}). To counteract this, BFT-Simulator compares execution traces of real deployments with traces from the simulation for validation (the authors are aware that this gives no strict guarantees for correctness)~\cite{wang2022tool}.

The last property deserves explanation: It seems desirable to also evaluate BFT protocols in an adverse environment, such as when a portion of nodes becomes faulty.
To our best knowledge, BFTSim only supports benign faults (i.e., faulty replicas staying silent~\cite{singh2008bft}), while the BFT Simulator from Wang et al.~\cite{wang2022tool} also supports some more sophisticated attacks (such as partition, adaptive and rushing attacks~\cite{wang2022tool}). 
The generic simulators and emulators which were not crafted for BFT research do not consider a global Byzantine attacker. 

Reproducibility and experimental controllability are important but seem to be provided by most if not all simulators and emulators so these properties are not used in our comparison.

\subsection*{\textbf{B} Reproduce our Results}

Our evaluation results can be reproduced. First, it is necessary to clone our toolchain repository and follow the setup instructions in the README file. For best compatibility, we recommend (and currently use) Ubuntu 20.04 LTS and Shadow v2.2 (the newest version as of time of writing) and Node version 16.3.0. If you want to simulate specific BFT protocols like HotStuff, Themis or BFT-SMaRt you will need to install their dependencies, too.

A series of experiments (like the "p1024" experiment row with increasing $n$) is specified in an experiments description file, in yaml format. The structure and description is easy to understand. For instance, an experiment for Hotstuff with 128 replicas looks like this:

\begin{minted}[
    gobble=0,
    frame=single,
    style=friendly,
    fontsize=\footnotesize,
    escapeinside=||
  ]{yaml} 
    protocolName: hotstuff
    protocolConnectorPath: ./connectors/hotstuff.js
    experiments:
      - 128rep: 
          misc:
            duration: 30 s
            parallelism: 16
            useShortestPath: false
          network:
            bandwidthUp: 10 Gibits 
            bandwidthDown: 10 Gibits 
            latency:
             uniform: true
             replicas: 1000 us
             clients: 1000 us
          replica: 
         |\colorbox{Yellow}{  \textbf{replicas}: 128 }|
            blockSize: 400
            replySize: 1024
          client:
            clients: 16
            numberOfHosts: 2
            startTime: 0 s
            outStandingPerClient: 175
            requestSize: 1024
\end{minted}

We put a folder called \texttt{examples/serial22} on the repository that contains all experiments description files that we used for the evaluation section of this paper. We also provide all of our data sets as a reference. 
After following the README, you can run simulations by typing in a shell:

\begin{minted}[
    gobble=0,
    frame=single,
    style=friendly,
    fontsize=\footnotesize,
    escapeinside=||
  ]{bash} 
npm run simulation -- examples/hotstuff/hs3-aws.yaml
\end{minted}

This will create a data set called \texttt{results.csv} in your experiments  directory.

\subsection*{\textbf{C} Create your own Experiments}

It is also easy to setup experiments that simulate AWS deployments. Here is an example:

\begin{minted}[
    gobble=0,
    frame=single,
    style=friendly,
    fontsize=\footnotesize
  ]{yaml} 
   protocolName: bftsmart
   protocolConnectorPath: ./connectors/bftsmart.js
   experiments:
     - 4replicasAWS:
       misc:
        duration: 1200 s
        parallelism: 16
        useShortestPath: false
       network:
        bandwidthUp: 1 Gbit
        bandwidthDown: 1 Gbit
        latency:
          uniform: false
          replicas: ['us-west-1': 1, 'eu-west-1': 1,
          'sa-east-1':1, 'ap-southeast-2':1] 
          clients: ['us-west-1': 1]
       replica:
        replicas: 4
        blockSize: 100
        replicaInterval: 100
        replySize: 0
        stateSize: 0
        context: false
        replicaSig: nosig
       client:
        clients: 1
        threadsPerClient: 1
        opPerClient: 2000
        requestSize: 0
        startTime: 30 s
        clientInterval: 0
        readOnly: false
        verbose: true
        clientSig: nosig
\end{minted}
The geographic distribution of replicas can be simply specified by passing a mapping of regions and the number of replicas to be placed in the respective region such as:

\begin{minted}[
    gobble=0,
    frame=single,
    style=friendly,
    fontsize=\footnotesize,
    escapeinside=||
  ]{bash} 
[ 'us-west-1': 1, 'eu-west-1': 1, 'sa-east-1':1, ..]
\end{minted}

\begin{listing}
\begin{minted}[
    gobble=0,
    frame=single,
    style=friendly,
    fontsize=\footnotesize,
        escapeinside=||
  ]{js} 
const processName = 'my-app'; // replace with the name of 
                             // your protocol app!

function getProcessName() { return processName; }

function getExecutionDir() {...}

function getExperimentsOutputDirectory() {...}

async function build(replicaSettings, 
                     clientSettings, 
                     log) {...} // Mandatory

async function configure(replicaSettings, 
                         clientSettings, 
                         log) {...} // Mandatory

async function getStats(log) {...} // Optional, called 
                                  // after simulation

module.exports = {build, configure, getStats, 
getProcessName, getExecutionDir,
getExperimentsOutputDirectory};
\end{minted}
\caption{Stubs of a BFT protocol connector.}
\label{protocol-connector}
\end{listing}

We think it is easy to adapt our experiment description files. However, some care needs to be taken. In the following, we want to share some insights we made:

\textit{Duration}: This is one of the most important parameters because it has a big impact on resource consumption of simulations. It is better to use short durations to keep the overall simulation time short.

\textit{Parallelism}: Make sure to set this parameter to utilize multi-core systems and speed up simulations.

\textit{BlockSize and OutStandingRequestsPerClient}: In HotStuff, if the number of in-flight requests is too low to fill the blocks, then HotStuff replicas just wait for more requests (which wont arrive). In this case the system halts and the simulation fast-forwards and terminates. It is better to overestimate the number of inflight requests for HotStuff.

Moreover, we recommend reading the documentation of parameters\footnote{\url{https://shadow.github.io/docs/guide/shadow_config_spec.html}}
and we also recommend to use \texttt{tmux} to run your simulations in the background.

\subsection*{\textbf{D} Simulate your own BFT Protocol Implementation}

You can implement your own BFT protocol connector by implementing the stubs \texttt{build, configure, getExecutionDir}
\texttt{getProcessName} and \texttt{getExperimentsOutputDirectory} as shown in Listing~\ref{protocol-connector}. Your connector can optionally also implement a method \texttt{getStats} to automatically parse logs and put the results in a comma separated value file.
Experiment description files have to adhere to a certain format:

\begin{minted}[
    gobble=0,
    frame=single,
    style=friendly,
    fontsize=\footnotesize
  ]{yaml} 
protocolName: #name of your protocol
protocolConnectorPath: #path to your connector
experiments: #Array describing the experiments
   -exp1: #description of an experiment
        misc: #miscelleanous settings 
            duration: #duration of the experiment
            useShortestPath: #default is false
            parallelism: # multi-core awareness
        network:
            bandwidthUp: ..
            bandwidthDown: ..
            latency:
                uniform: (true|false)
               # if true:
                    replicas: #inter-replica 
                    #latency ex. 1000 us
                    clients: #client-replica 
                    #latency ex. 1000 us
               #  else:
                    replicas: #Array  describing 
                    #AWS hosts format region: host
                    clients: #Array  describing 
                    #AWS hosts format region: host 
                    #OR: a uniform client-replica latency
        replica: 
            #This is for protocol- and replica-specific 
            # configs; will be passed to your connector.
        client:
            #This is for protocol- and client-specific
            # configs; will be passed to your connector.
\end{minted}
You may leverage the .env file to make your experiment description files and your connectors more concise.

\subsection*{E Determinism in Phantom}

Throughout the simulation, Phantom preserves determinism: It employs a pseudo-random generator, which is seeded from a configuration file to emulate all randomness needed during simulation, in particular the emulation of \texttt{getrandom} or reads of \texttt{/dev/*random}.
Each Phantom worker only allows a single thread of execution across all processes it manages so that each of the remaining managed processes/threads are idle, thus preventing concurrent access of managed processes’ memory~\cite{jansen2022co}.
%We briefly display the architecture in Figure~\ref{fig:phantom:architecture}, for more details we refer the reader to the Phantom paper~\cite{jansen2022co}.

\end{document}